\documentclass[useAMS,usenatbib,referee]{mn2e}
\usepackage{deluxetable}
\usepackage{comment}
\usepackage{ifthen}
\usepackage{amsmath}
\usepackage{amsfonts}
\usepackage{amssymb}
\usepackage{graphicx}
\usepackage{wasysym}
\usepackage{subfigure}
\usepackage{lscape}

\newcommand{\forloop}[5][1]%
{%
\setcounter{#2}{#3}%
\ifthenelse{#4}%
	{%
	#5%
	\addtocounter{#2}{#1}%
	\forloop[#1]{#2}{\value{#2}}{#4}{#5}%
	}%
	{%
	}%
}%

\newcommand{\ctbd}[1]{}

\newcommand{\lcs}{light curves}

\newcommand{\rsun}{\ensuremath{R_\odot}}

\newcommand{\rearth}{\ensuremath{R_\oplus}}

\title[Testing the Titius-Bode law predictions for {\em Kepler} multi-planet systems]{Testing the Titius-Bode law predictions for {\em Kepler} multi-planet systems}
\author[Chelsea X. Huang and G\'asp\'ar \'A. Bakos]{Chelsea X.Huang$^{1}$\thanks{E-mail:xuhuang@princeton.edu;},
G\'asp\'ar \'A. Bakos$^{1,2,3}$ \\
$^{1}$ Department of Astrophysical Sciences, Princeton University, NJ 08544, USA. \\
$^{2}$ Alfred P. Sloan Research Fellow. \\
$^{3}$ Packard Fellow. 
}

\begin{document}

\maketitle

\begin{abstract}

We use three and half years of {\em Kepler} Long Cadence data to search for 
the 97 predicted planets of Bovaird \& Lineweaver (2013) in 56 of 
the multi-planet systems, based on a general Titius-Bode relation. 
Our search yields null results in the majority of systems. We 
detect five planetary candidates around their predicted periods. 
We also find an additional transit signal 
beyond those predicted in these systems. 
We discuss the possibility 
that the remaining predicted planets are not detected in the {\em Kepler} 
data due to their non-coplanarity or small sizes. We find that the 
detection rate is beyond the lower boundary of the expected number 
of detections, which indicates that the prediction power of the TB relation 
in general extra solar planetary systems is questionable. Our analysis of 
the distribution of the adjacent period ratios of the systems suggests 
that the general Titius-Bode relation may over-predict the presence of 
planet pairs near the 3:2 resonance.
\end{abstract}
\section{Introduction}
\label{sec:intro}
The {\em Kepler} mission \citep{Borucki:2010, Batalha:2013} has dramatically 
increased our knowledge about the architecture of extrasolar planetary 
systems. More than 3600 planetary candidates have been announced in the 
$\sim 4$ yrs time span of the mission. A significant fraction of these 
planet candidates reside in multiple planet systems \citep{Latham:2011}. 
Based on the most recent candidate list, \citet{Batalha:2013} reports 
that $20\%$ of the planet hosting stars have multiple planet candidates 
in the same system.

The multiple planetary systems discovered by {\em Kepler} show a variety of 
structures. Various authors 
(e.g. \citet{Lissauer:2011,Fabrycky:2012,Steffen:2013}) generally agree 
on the following features of the {\em Kepler} multiple systems: 
a) the majority of the multiple systems consist of several Neptune 
or super-earth size planets packed within 100 day period orbits;
b) the multiple systems are highly coplanar;
c) unlike the Solar System, most of the planet candidates are not 
in or near mean-motion resonances (MMRs); however, MMRs are clearly 
preferred to what would be expected from a random distribution of 
period ratios \citep{Petrovich:2013}. To our best knowledge, these features are not fully reproduced 
by current theories.

Many centuries ago, the architecture of our Solar System was proposed to 
follow a relation, which describes the semi-major axes of Solar 
System planets in a logarithmic form as a function of their 
sequences in the system. 
\begin{equation}
a {\rm(AU)} = 0.4 + 0.3 \times 2^n,\,\,n=-\infty,0,1,2 ... 
\end{equation} 
This Titius-Bode relation (hereafter TB) 
successfully predicted the existence of the Asteroid Belt 
and Uranus, based on the knowledge of the orbits of other planets (Mercury 
through Saturn) in the Solar System. 

The physical origin, if any, of the TB relation is not well understood.  
The explanations that have been proposed to reproduce this phenomenon 
can be grouped in the following categories: 
a) there are physical processes that directly lead to the TB relation, for 
instance, dynamical instabilities in the protoplanetary disk \citep{Li:1995}, 
gravitational interactions between planetesimals \citep{Laskar:2000}, 
or long term dynamical instabilities \citep{Hills:1970};
b) it is a statistical result of some physical requirements in planetary 
systems, such as the radius exclusion law 
based on stability criteria \citep{Hayes:1998};
c) it is a presentation of other spacing laws, for example, the capture 
in resonances between the mean motion of the planets \citep{Patterson:1987}. 

\citet{Bovaird:2013} (hereafter BL13) tested the applicability of TB 
relation on all of the known high multiplicity extrasolar planetary 
systems (those with four or more planets around the same hosting star).
Using the systems that they considered to be complete by stability 
requirements,  they found that rather than following the exact same 
TB relation as our Solar System, 94\% of these complete systems favor 
a more general logarithmic spacing rule (the general TB relation).
For a system with N planets, the general 2-parameter TB relation 
(which, for simplicity, we no longer distinguish from the Solar System law 
in the rest of text) can be 
formalized as:
\begin{equation}
{\mathrm log}\,P_n = {\mathrm log}\,P_0 + n* {\mathrm log}\,\alpha,\,\,\, n = 0,1,2,...N-1.
\end{equation}
They further inferred that all the high multiplicity systems prefer 
to adhere to this general TB relation: if a system does not 
follow this relation as tight as the Solar System, there is a high possibility 
that one or more planets are not detected in this system. 
They predicted that $\sim32\%$ of the {\em Kepler} high multiplicity systems should 
host additional planets within the orbit of the outermost detected planets.

It is possible that these planets are missed from the {\em Kepler} sample due to 
their small sizes, that they are simply not transiting due to slightly 
different inclination respect to the rest of disk, or they are missing due to 
the incompleteness of the {\em Kepler} pipeline. In this work, we carried out a 
careful search for the additional planets in these {\em Kepler} systems to test the 
predictions in BL13.

We use the Quarter 1-15 {\em Kepler} long cadence data in this analysis. We 
detected five planetary candidates around the predicted periods, 
found one additional transit signal 
that were not predicted in these systems. We did not find majority of the 
predicted systems. We describe our sample and restate the predictions from 
BL13 in Section \S 2. We then introduce our method in Section \S 3. 
Finally, we present our result and discuss the detection bias and 
indications in Section \S 4.

\begin{figure}
\includegraphics[width=\linewidth]{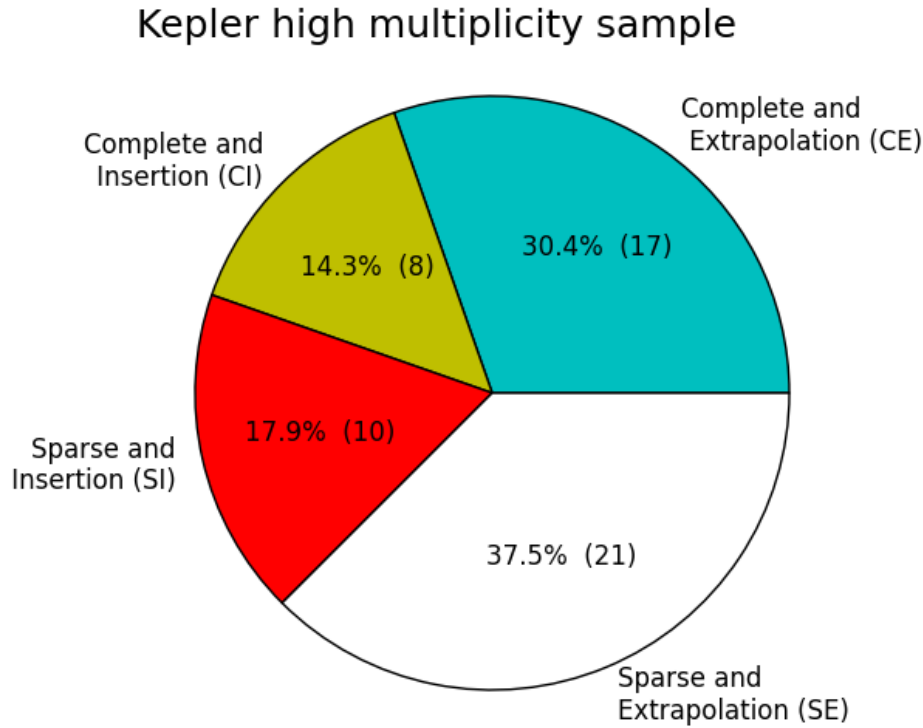}
\caption{
We show the distribution of {\em Kepler} high multiplicity systems 
in the four subcategories based on BL13. The four subcategories are 
``complete and insertion (CI)", ``complete and extrapolation (CE)", 
``sparse and insertion (SI)", ``sparse and extrapolation (SE)". The 
completeness of a system is measured by the spacing between two 
neighboring planets. The insertion condition is such that if a system 
has fitting statistic with the TB relation worse than the Solar System, 
planet insertions are made to the system.
\label{fig:pie}
}
\end{figure}

\begin{figure*}
\includegraphics[width=\linewidth]{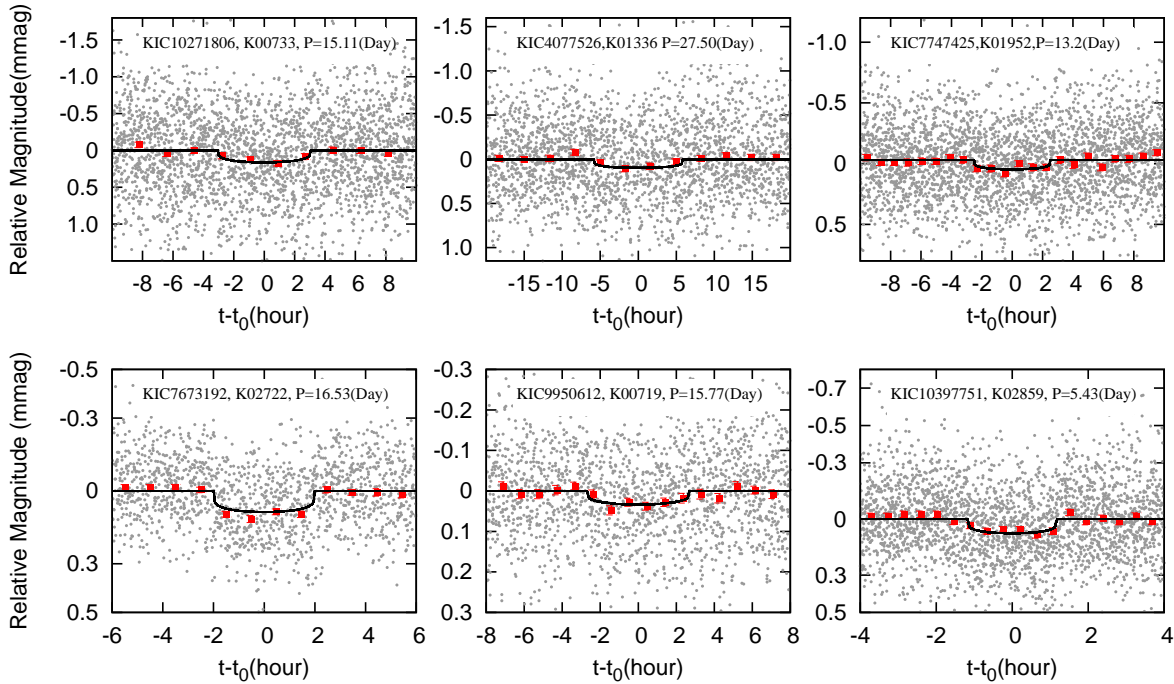}
\caption{
Folded light curves around the center of transits $t_0$ of the newly 
discovered planetary candidates from {\em Kepler} Q1-Q15 long cadence data. 
The unbinned data is shown in grey dots, 
while the binned data is shown in red squares with the error bars 
representing the uncertainties in binned average. Super-imposed are 
the best fitted models of the transits. 
We note that the y axis scales vary between subfigures. 
\label{fig:transits}
}
\end{figure*}


\section {The {\em Kepler} sample}

BL13 predicted the existence of 141 additional exoplanets in 68 
multiple-exoplanet systems.  
60 of these systems were discovered by the {\em Kepler} 
mission \footnote{See Kepler Object of Interest-Cumulative: http://exoplanetarchive.ipac.caltech.edu/cgi-bin/ExoTables/nph-exotbls?dataset=cumulative, also 
known as the 2013 Kepler catalog}. The remaining 8 are non-transiting 
planets discovered by radial velocity searches, which we do not include 
in this analysis. They suggested that altogether 117 planets were 
missing in these {\em Kepler} systems. 

We excluded four systems in our analysis as follows. 
BL13 reported analysis for Kepler-62 as well as for KOI-701, but these 
identifications refer to the same planetary systems. Thus, we exclude 
KOI-701 from our analysis. The period of KIC8280511 (KOI-1151.01) is more 
likely double the period quoted by the {\em Kepler} catalog. With a 
revised period of 10.435374 days \footnote{The period of this planet has 
been corrected in the 2014 {\em Kepler} catalog.}, this system adheres to 
the TB relation tighter than the Solar System without the need of inserting 
additional planets. The other two excluded systems were eliminated, 
because we could not establish the validity of their fourth planetary 
candidate. To be more specific, for KIC3447722 (KOI-1198), we were not able to 
recover the fourth transit (KOI-1198.04, with a period of 1.008620 days) 
in our analysis. For KIC8478994 (KOI-245, also known as Kepler-37), the 
fourth transit listed in the {\em Kepler} catalog 
(KOI-245.04, $P=51.198800$ days) was not detected by our pipeline, while 
the discovery paper of Kepler-37b \citep{Barclay:2013} also stated only 
three planets in this system. 
 
The predictions from BL13 can be grouped into two categories: the ``inserted" 
planets and the ``extrapolated" planets. Among all the {\em Kepler} 
high multiplicity systems (those with four or more planets around the same 
host star), BL13 found that 38 of them fit the TB relation comparable 
to or tighter than our Solar System. The tightness of the fitting is 
determined by the $\chi^2_\nu$ ($\frac{\chi^2}{\rm d.o.f.}$) derived from 
the fitting based on the Equation 4 of BL13. 
They predicted 41 ``inserted" planets for the remaining 18 systems, arguing 
that the insertion of additional planets will make a better fit for the 
TB relation. Additionally, they hinted on the possible existence of 
one ``extrapolated" planet for each system (56 in total) using the best 
fit of the TB relation accounting for the ``inserted" planets. 

On the other hand, it is also important to divide these 
systems according to their completeness. 
The completeness of a planetary system is measured by the dynamical 
spacing $\Delta$, defined as $(a_2-a_1)/R_{\rm H}$, where $a_1$ 
and $a_2$ are the semi-major axes of the neighboring pair of planets. We 
adopt the Hill radius $R_{\rm H}$ defined in \citet{Chambers:1996}, 
\begin{equation}
R_{\rm H} = [(m_1+m_2)/3M_\odot]^{1/3}[(a_1+a_2)/2], 
\end{equation}
where $m_1$ and $m_2$ are the planetary masses. The masses of the planets 
are estimated by a mass-radius relation conversion 
$M_p=(R_p/R_{\oplus})^{2.06}M_{\oplus}$ \citep{Lissauer:2011}. 
BL13 isolated 25 planetary systems as their ``most complete" sample using 
the dynamical spacing criteria 
\footnote{The dynamical spacing criteria for completeness of a planetary 
system are a) all adjacent planet pairs have $\Delta$ values smaller 
than 10 if an additional planet is between each pair. b) at least two 
adjacent planet pairs in the system have $\Delta$ values smaller than 10 if 
additional planet pairs are inserted.}. 

Therefore, we further divided the {\em Kepler} high multiplicity samples, yielding 
altogether four non-overlapping categories (see Figure \ref{fig:pie}) using 
the above two criteria from BL13. 
Among the 25 most complete systems, there are 17 systems that originally 
have a better fit to the TB relation than the Solar System (thus only 
one ``extrapolated" prediction is made in each of the systems) (we call 
this category ``CE" for short in the text, with ``C" stands for complete 
and ``E" stands for extrapolated); the remaining 8 systems are ``most" 
complete as defined by BL13, but require insertion to fit better than 
solar (we call it ``CI" for short, 
where ``I" stands for insertion). Another 10 systems are sparse and 
need insertion to fit better than solar (``SI" for short, where ``S" stands 
for sparse); 21 systems are sparse but have an original fit better than 
solar (``SE" for short).
These 56 systems were analyzed and searched for the 97 predicted planets.

\section{Analysis}

We revisited all 56 systems with the Q1 to Q15 long cadence {\em Kepler} 
public data (with more than 1000 days total time span), looking for 
additional periodic signals.
First we validated the detectability of all the known {\em Kepler} planetary 
candidates in these systems. We applied the Box Least Square 
Fitting (BLS) algorithm to the pre-filtered \lcs. The pre-filtering processes 
included the removal of bad data points from the {\em Kepler} raw data 
(the ${\rm SAP\_FLUX}$), the correction of safe modes and tweaks in 
the \lcs, and the filtering of the systematic effects and stellar 
variations by a set of cosine functions with a minimum period of 1 day. 
We refer to \citet{Huang:2013} for more details 
about our pre-filtering technique. We then cross referenced our 
detected signals with the reported periods and epochs from the 
2013 {\em Kepler} catalog. 

All the known transit signals in these systems have signal to noise (SNRs) 
higher than 12 and dip significance (DSPs) higher than 8. 
These known transits from the systems were then removed from the 
raw (${\rm SAP\_FLUX}$) \lcs\ with a window function 1.1 times wider than the 
reported transit duration from {\em Kepler} catalog. The transits with high 
Transit Timing Variations (TTVs) (such as KOI-250 \citep{Steffen:2012a} 
and KOI-904 \citep{Steffen:2012b}) were removed by visual inspection of
the \lcs. 

After the removal of all the known signals, the raw \lcs\ were then 
cleaned by the same pre-filtering technique as above. 
We ran the BLS again on these \lcs\ to ensure that the known signals have 
been fully removed. We selected the BLS peaks with the same 
threshold above. We also visually examined local maximums in the 
BLS spectrum within the predicted range of BL13 (regardless of their 
signal to noise). The selected peaks were then checked against our 
standard vetting procedures 
in \citet{Huang:2013}. Five new periodic planetary candidates 
were detected that passed our vetting procedures. 
We then derived the best fit parameters and their error bars of the vetted 
planetary candidates using the Markov Chain Monte Carlo algorithm. 

\begin{figure}
\includegraphics[width=\linewidth]{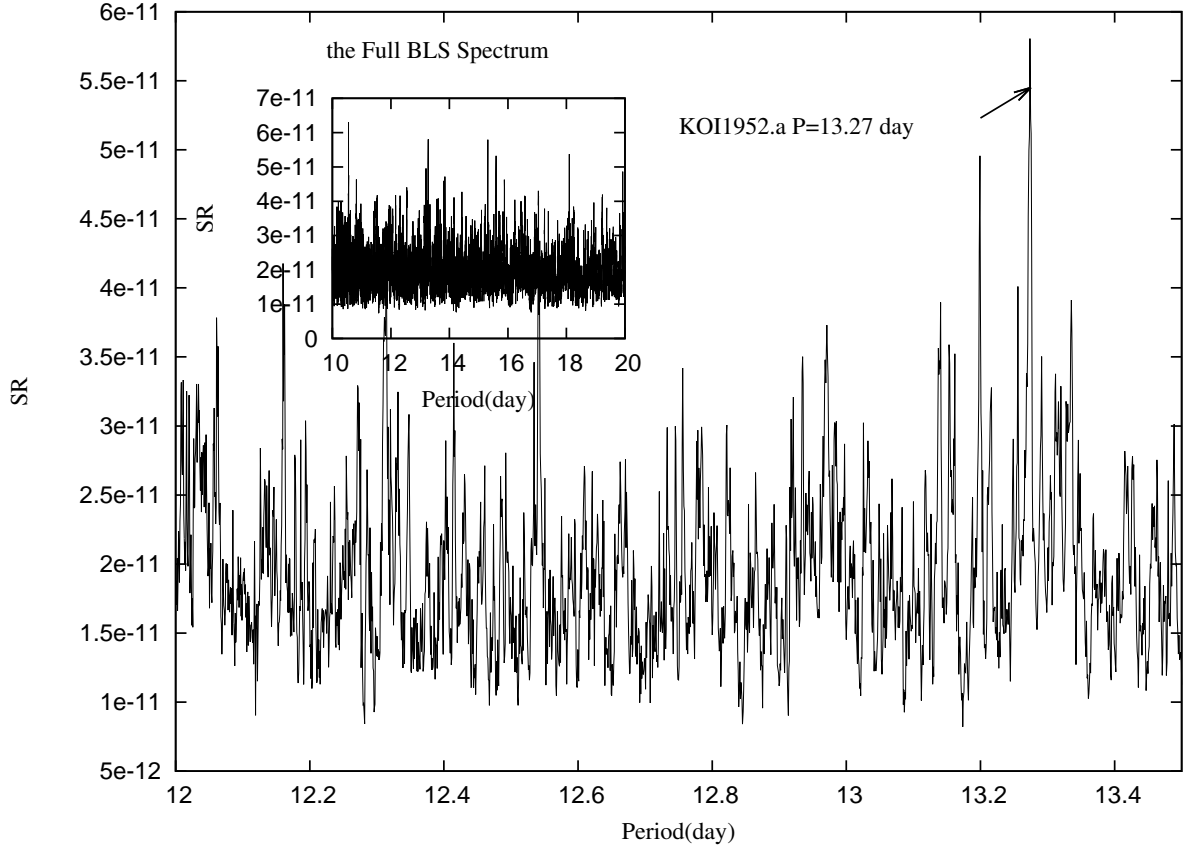}
\caption{
The BLS spectrum of KOI-1952 after the known transit signals removed. 
We zoomed in around the predicted period of the transit signal. 
The full spectrum is shown in the small window. All the other peaks of 
comparable height with the detected transit signal have much lower dip 
significance.
\label{fig:bls}
}
\end{figure}

We show the BLS spectrum of KOI-1952 in Figure \ref{fig:bls}. 
The detected period has a signal to noise (SNR) of 13.7 for the 
spectrum peak and a dip significance (DSP) of 9.0. There are three more 
peaks with comparable SNR in the spectrum, but their DSPs are all 
less than 8.

\begin{figure}
\includegraphics[width=\linewidth]{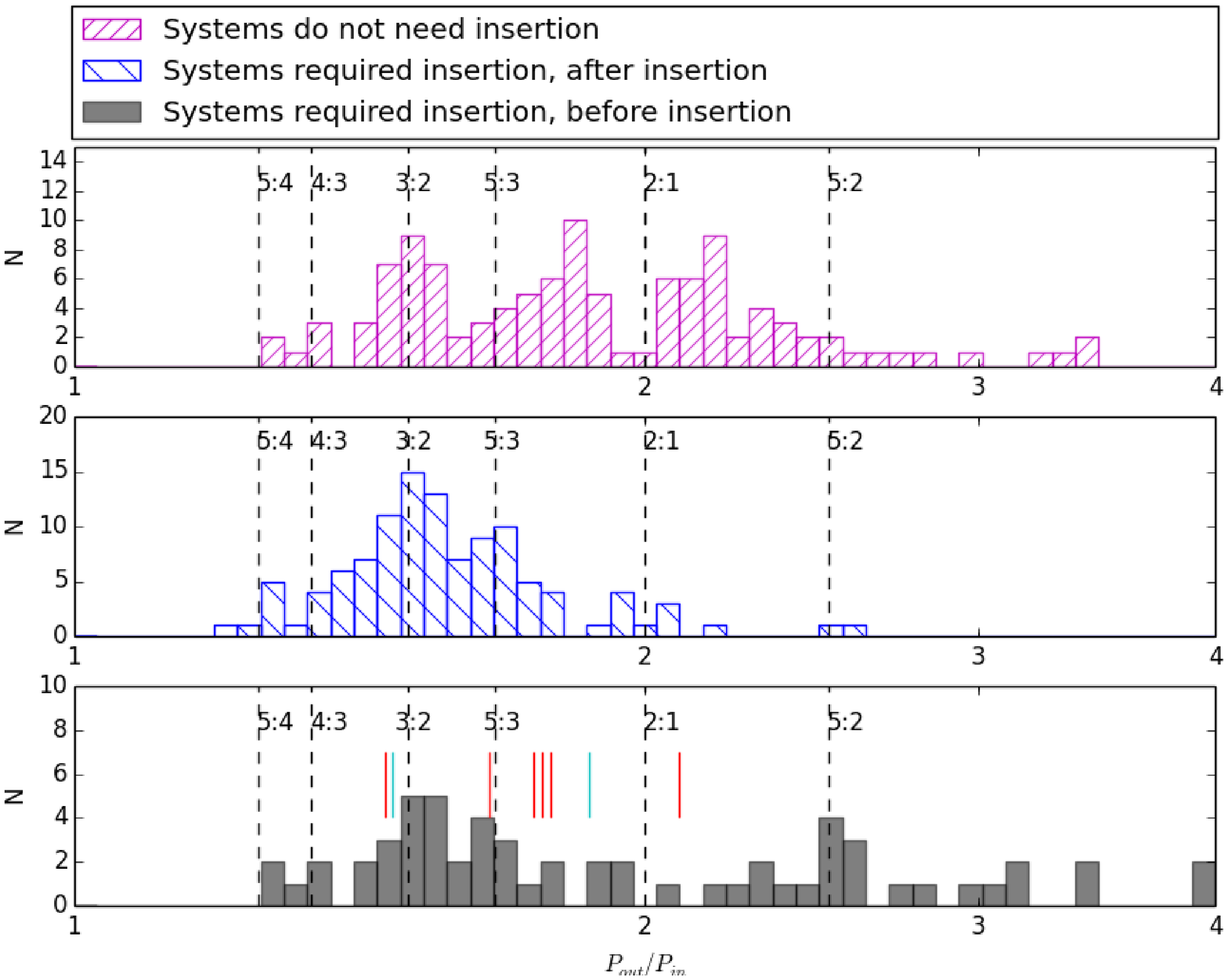}
\caption{
Histogram of period ratio between the neighboring planet pairs in 
{\em Kepler} high multiplicity samples. We present the systems that 
do not need insertion in magenta and hatched with `/'. 
The systems that required insertion, with all the inserted planets taken into 
account are shown in blue and hatched with `\textbackslash{}'.
We also shown the original (before insertion) period ratio distribution 
of systems needing insertion in grey. We marked out the position of the 
new discoveries in the bottom panel with vertical bars (red for ``inserted" 
planets and cyan for ``extrapolated" planets.) for references. 
\label{fig:ratio}
}
\end{figure}

\section{Results and Discussion}

\subsection{Summary of Detections}
In this search among the 56 {\em Kepler} high multiplicity systems, using 
SNR $>12$ and DSP $>8$ criteria, and visually inspecting the BLS spectrum 
in the predicted period range, we have not found the majority of the 
predicted planets from BL13. We found five of the predicted signals, 
and one periodic transit 
that was not predicted. 
We present the newly discovered transit signals in 
Figure \ref{fig:transits}. The transit parameters are listed against the 
predicted periods in Table \ref{tab:parameters}. We also list the known 
planetary candidates from {\em Kepler} team and their detection SNR and DSP in 
Table \ref{tab:parameters} as a comparison. Most of the detected signals 
indicate planets with size smaller than $2$\rearth. The detection SNR and 
DSP is on the low side of the overall population of {\em Kepler} candidates. 
However, the factor that they reside in multiple planet systems enhance 
their chances to be real planets \citep{Lissauer:2012}. 

Details specific to each of the systems are spelled at below.

{\bf KOI-719} was predicted to host three ``inserted" planets and one 
``extrapolated" planet. We found a planetary candidate with period of 
15.77 days around the predicted period $14\pm2$ days. We do not see 
the rest of predicted planets in this system. Since KOI-719 is a small 
star ($R_{\rm star} = 0.66$\rsun), this candidate is extremely small, with best 
fit radius similar to Jupiter's moon Ganymede, $\sim0.42$\rearth. This 
new signal is around the 7:4 resonance of both its inner neighbor 
and its outer neighbor, reduces the system $\chi^2_\nu$ from 
1.35 to 0.5. 

{\bf KOI-1336} was predicted to host a $6.8\pm3$ day 
and another $26\pm3$ day ``inserted" planet. We recovered the latter at 
27.5 day period. This signal is outside the 5:3 resonance with the inner planet 
(KOI-1336.02), and near the 2:3 resonance with the outer planet (KOI-1336.03). 
The $\chi^2_\nu$ of this system slightly increased with the insertion 
of this signal (from 1.06 to 1.35). If the $6.8$ day planet exists (which we 
did not see in this search), this value would drop to 0.25.
We failed to recover the ``extrapolated" $61\pm6$ day signal.

{\bf KOI-1952} was predicted to host two additional ``inserted" planets, 
with periods around $13\pm2$ days and $19\pm2$ days and an ``extrapolated" 
planet with period of $65\pm7$ days. The detected signal 
matches well with the first prediction, and has a period of 13.3 days. This 
period is near the 3:2 resonance of KOI-1952.01, and near the 1:2 resonance 
of KOI-1952.02. We did not find any significant signal around the 19 day 
period. With the insertion of this transit signal, the $\chi^2_\nu$ 
of this system decreased from 3.25 to 0.64, following a tighter TB relation.  

{\bf KOI-2722} was predicted to host an ``extrapolated" $16.8\pm1$ day 
transit. With the Q1-Q15 \lcs, we recovered a new signal at a period of 
$16.5$ days, which is also reported by {\em Kepler} team in the recent 
accumulated catalog. This outer planetary candidate is near the 3:2 
resonance with the previous outmost planetary candidates in the system. 
The next outer planetary candidate, predicted by BL13 around period 
of $24\pm2$ days, was not identified in our analysis.

{\bf KOI-2859} was predicted to host an ``inserted" planet with period 
around $2.41\pm0.1$ days and an ``extrapolated" planet with period 
around $5.2\pm0.3$ days. We only found the $5.43$ day ``extrapolated" 
signal. It is not in tight resonances with any of the other planets 
in the system. The $\chi^2_\nu$ is improved to be 1.07.  

We also have some detections that do not match the predictions.
We found a new ``inserted" planetary candidate (with P$=$15.1 days) 
in the {\bf KOI-733} system, which was predicted to host an ``extrapolated" 
planet with a period of 36 days. Although this system original fits 
better than solar, it was clarified as dynamically unpacked. Adding 
in this planet in the system will introduce the 4:3 
resonance to the inner planet and the 4:5 resonance to the outer planet. 
However, this possible insertion will make the TB relation less tight; 
$\chi^2_\nu$ increased from 0.22 to 4.06.

We put these discoveries in context of the four subcategories we stated 
in Section \S2 (Table \ref{tab:expect}). We found 3 ``inserted" and 
1 ``extrapolated" planetary candidates around the predicted periods in 
the 10 ``sparse and insertion (SI)" systems, and 1 ``extrapolated" planetary 
candidate around the predicted period in the 17 ``complete and 
extrapolated (CE)" systems. We did not find any transit signal within 
the error-bar of the predicted periods of the 8 ``complete and 
insertion (CI)" systems and 21 ``sparse and extrapolated (SE)" systems.

\subsection{Detection bias analysis}

We considered some possible explanations for the discrepancy between the 
predicted planets and the observed numbers. 
We first look at the transit probabilities of the predicted planets. 
The maximum inclined angle for a planet to transit a solar 
type star can be expressed as a function of its period:
\begin{equation}
i_{\rm max} \approx 1.06^{\circ}(\frac{P}{50\,{\rm day}})^{-2/3}.  
\end{equation}
\citet{Fabrycky:2012} found that the typical mutual inclinations of 
{\em Kepler} multiple systems lie firmly in the range $1.0^\circ$-$2.3^\circ$, 
most consistent with a Rayleigh distribution peaks at $1.8^\circ$. 
We consider a Rayleigh distribution of planet inclinations constrained 
by \citet{Fabrycky:2012}, with the Rayleigh width\footnote{Here 
$1.0^\circ$ and $2.3^\circ$ are 
the 3-$\sigma$ lower and upper limit for the possible range of 
Rayleigh width. See Figure 6 of Fabrycky et al (2012).} 
$\sigma_i = 1.8_{-0.8}^{+0.5}\,^\circ$, and assume all the stars to 
have solar radii. 
If we assume the mean planes of the known planets in every 
system lie in the plan including the line of sight and the stellar mid-disk, 
the expected number of planets to transit can be expressed with 
\begin{equation}
\sum\limits_{i=1}^n \frac{C_0^{i_{\rm max}}}{C_{0}^{90}} \approx \sum\limits_{i=1}^n C_0^{i_{\rm max}} = \sum\limits_{i=1}^n (1-\exp{[-\frac{1.06^2}{2\sigma_i^2}(\frac{P_i}{50 {\rm day}})^{-4/3}]}).
\end{equation}
We use $C_0^{i}$ to note the cumulative distribution of a Raleigh 
distribution from 0 to angle i. The maximum possible value of i is 90$^\circ$ 
due to geometric symmetric reason. 
$P_{i=1,n}$ is the discrete period distribution of the predicted planets 
taken from BL13. In reality, the assumption about the mean planes of the 
known planets lie in the middle disk of the stars are not strictly true. 
Therefore, our estimation of the expected number of planets that transit 
shall be slightly smaller than the true expected number. 

We found that $11_{-3}^{+12}$ out of 56 of the ``extrapolated" planets are 
expected to transit their host stars, while $22_{-4}^{+10}$ out of 41 of the 
``inserted" planets are expected to transit their host stars. We estimated 
the upper limit here using $\sigma_i=2.3^{\circ}$, and the lower limit 
with $\sigma_i=1.0^{\circ}$. 
The ``extrapolated" planets usually have longer periods, which are less 
likely to be detected than the ``inserted" planets.
It would be interesting for future works to search for transit timing 
variations of the known planets in these systems to reveal potential 
non-transiting signals. 

The rest of the non-detected planets could also be missing due to their small 
sizes. \citet{Howard:2012} proposed that the planet size distribution follows 
a power-law for all the {\em Kepler} planetary candidates,
\begin{equation}
\frac{{\rm d}\,f(R)}{{\rm d}\,R} = 2.9R^{-2.92}. 
\label{eq:H12}
\end{equation}
The power-law distribution in the radius range above 2 \rearth\ has been 
confirmed by various authors \citep{Dong:2012, Petigura:2013a, Petigura:2013b}.
However, this estimation of planet occurrence rate becomes incomplete for 
$R_p<2$\rearth\ (\citet{Howard:2012}, Fig 5). \citet{Petigura:2013b} used 
an injection-recovery technique to produce the current best estimation 
of the occurrence rate for the small planets. 
We fit the distribution from \citet{Petigura:2013b} (Figure 3A) with 
$R_p<2.8$\rearth\ using a power-law distribution as Equation \ref{eq:P13} 
and then use it as our default radius distribution in the radius range 
$R_p<2$\rearth.

\begin{equation}
\frac{{\rm d}\,f(R)}{{\rm d}\,R} = kR^{-\alpha}, 
\label{eq:P13}
\end{equation}
where $k =0.30\pm0.013$, $\alpha=0.36^{+0.05}_{-0.07}$. 
The error-bar of the power-law index is estimated with a Monte-Carlo method, 
taking into account of the uncertainties in the occurrence rate presented by 
\citet{Petigura:2013b}. However, we caution the reader that 
this fitting is based on only three data bins with planet radius in the 
range of 1-2.8 \rearth. Therefore the tightness of these error-bars do not 
well represent the statistical confidence of the power-law indexes. Moreover, 
the exact turn over point of the distribution is uncertain, likely to lie 
between $2-2.8$\rearth\ based on the study of \citet{Petigura:2013b}. 
We adopt $2$\rearth\ as the transition point. We assume here for planets 
with size larger than the transition point, the planets follow a distribution 
described by Eq.\ref{eq:H12}, while smaller planets follow a distribution 
described by Eq.\ref{eq:P13}. The two distributions are scaled such that the 
${\rm d}f(R)/{\rm d}R$ value at 2\rearth\ is the same.

For each planet, we take into account the stellar properties and 
the light curve precision in the correction and assume completeness in the 
detection for all the planets with signal 8 times than the noise level 
in phase space. 
The minimum size of planet we assume to be completely detected for 
each light curve can be expressed as 
\begin{displaymath}
r_{\rm min}=r_{\rm star}[8\sigma_{\rm lc}(\frac{1400 \rm day}{P})^{1/2}]^{1/2}. 
\end{displaymath}
Therefore the detection probability for each transit is 
\begin{equation}
Pr({\rm Detection})
=(1-\exp{[-\frac{1.06^2}{2\sigma_i^2}(\frac{P_i}{50 {\rm day}})^{-4/3}]})
\times(\int_0^{r_{\rm min}}\frac{{\rm d}f(R)}{{\rm d}R}\,dR).
\end{equation}
We found $4.2_{-1.2}^{+3.5}$ extrapolated planets and $10.7_{-1.8}^{+3.6}$ 
inserted planets are expected in this detection. 
We remind the readers the error-bars 
here represent the 3$-\sigma$ limit from the Rayleigh distribution. 
The number of detections we found for both the ``extrapolated" (2 detections) 
and the ``inserted" planets (3 detections) are far beyond the lower limit of 
the predicted expectations.  

To explore the effect of a different size distribution, we also report 
the expected number of detections with the power-law distribution 
$\frac{{\rm d}\,f(R)}{{\rm d}\, R} = kR^{-1}$ (constant occurrence rate in log 
space) in this radius range ($R<2$\rearth), as suggested by 
\citet{Petigura:2013a}.
To avoid the singularity of integration with the plateau distribution at 
size 0, we set the lower limit of the planet size distribution to be 
$0.1$\rearth. One could of course argue that all the missing planets might 
as well be even smaller bodies than this, but it is out of the scope of 
this paper to discuss the region beyond our understandings.    
We found $2.7_{-0.7}^{+2.2}$ extrapolated planets and $7.3_{-1.2}^{+2.3}$ 
inserted planets are expected under this assumption.

We summarize the expected number of detection in the context of the four 
category of systems in Table \ref{tab:expect}, and compare them with 
detections. Generally speaking, our number of detections are below the 
expected lower limit of detections in most of the cases. 
Predictions in the ``CI" systems are ruled out at higher than 
$20\sigma$ level. Although BL13 claim majority (94$\%$) of the most 
complete systems tend to follow a tight TB relation, our findings 
indicate that those most complete systems that do not follow a tight TB 
relation could not be explained by missing planets. 

\subsection{Period ratio distributions}
It is difficult to validate most of the predictions from the TB relations 
due to the uncertainty in the size distribution for small planets. 
However, we can use the period ratio distribution to examine the predicted 
planet population further, since there is no known strong bias against 
detections of planets in particular low order period ratios. 
\citet{Steffen:2013} pointed out that there could be detection bias 
against adjacent period ratios that are larger than 5:1 or 6:1. In this 
analysis, we only focus on the region where adjacent period ratios are 
smaller than 3:1. We expect that the systems that do not need insertion 
(``CE" and ``SE") represent the underlying true period ratio 
distribution; while the planet systems needing insertion (``CI" and ``SI") 
should recover the same 
distribution after accounting for all the inserted planets. Our analysis 
here is independent of the number of predictions confirmed by our work.  

We test the above expectation with these two populations in 
Figure \ref{fig:ratio}. The distribution of neighboring period ratios 
for the ``CE" and ``SE" systems is shown with 
magenta; and the ``CI" and ``SI" systems with all their predicted inserted 
planets are shown in blue. For reference, we also plot in grey the 
distribution of ``CI" and ``SI" systems before the insertion. 
A striking feature is that the distribution of planetary systems after 
insertion looks significantly different from the systems that do not 
require insertion.
We also performed a two-sample Kolmogorov-Smirnof statistic 
on the two populations, obtaining a KS p-value of $3\times10^{-11}$.
It is unlikely that the ``CI" and ``SI" systems together with their 
inserted planets recovered the period ratio distribution suggested by the 
systems that do not require insertion.      
This indicates that some of the ``insertion" by BL13 might be problematic. 

By observing the detailed structures in the period ratio distribution, we 
found that systems following a tight TB relation (thus not needing 
insertion) is more likely to have period ratio around (wider to) the MMRs, 
such as 3:2, 5:3 and 2:1 resonances. Dynamic models 
(see \citet{Hansen:2013} and \citet{Petrovich:2013}) can also 
reproduce these features. We note that the tail of the 
distribution at period ratio high than 3 is not statistically 
significant, the chance of having non-detected planets in these 
few sparse systems are not negligible (if follow the same size and inclination 
distribution arguments in Section \S4.1).

On the other hand, the systems that do not follow well a TB relation 
before their insertion have less pronounced peaks near the above MMRs. 
Including all the predicted ``inserted" planets, such that the 
systems can be optimally fitted by a tight TB relation, will preferentially 
add in (or modify) the period ratios close to the lowest order 
resonance in the original system. Therefore, the revised period ratio 
distribution according to the TB relation over-predicts the clustering 
around the 3:2 MMR and under-predicts the 2:1 MMR. 
It is hard to explain that most of the missing planets belong to the 3:2 
MMR rather than the other MMRs by observation bias,  
unless there is a strong correlation between the period ratio of the 
adjacent planet and the size ratio of the adjacent planet, which is not 
seen in the previous study by \citet{Ciardi:2013}.

\section{Summary}
The Titius-Bode relation as been a recurrent theme in astronomy over the 
past two centuries. Investigations were all based on our Solar System - 
lacking other multiplanet systems. It is for the first time that validity 
of the TB relation can be tested on a statistically meaningful sample 
-- thanks to hundred of multi-planet discoveries made by the Kepler Space 
mission. By analyzing the multiple systems and assuming that a general TB 
relation holds, BL13 predicted 141 new planets. We found only 5 of the 
predicted planets in the Kepler data. Some of the planets may be not 
finding because of small size or a non-transiting inclination angle. 
Nevertheless, even after taking these observational biases into account, 
the number of detected planets is still fewer than the $3-\sigma$ lower 
limit of the expected planets from the prediction, hinting that it is 
questionable to apply such a law to all the extra-solar high 
multiplicity systems.  

\clearpage

\begin{landscape}
\begin{deluxetable}{lcclcccccc}
\centering
\tablewidth{0pc}
\tablecaption{Planet Parameters \label{tab:parameters}}
\tablehead{
\colhead{KIC} & \colhead{$\chi^2_\nu$\tablenotemark{a}} &\colhead{Origin\tablenotemark{b}} &\colhead{KOI} & \colhead{Period} & \colhead{Epoch(BJD-2454000)} & \colhead{$R_p/R_{\star}$} &\colhead{$R_p/\rearth$} & \colhead{SNR} & \colhead{DSP}
}
\startdata
9950612 & 1.35    & O & K00719.01 & 9.03    &   1004.014 & 0.023   & 1.64 & 1048    &  59 \\
 &     & O & K00719.02 & 28.12    &   979.913 & 0.0123   & 0.88 & 77    & 18 \\
 &     & O & K00719.03 & 45.90    &   999.538 & 0.0165   & 1.18 &  184   & 25 \\
 &     & O & K00719.04 & 4.159    &   966.784 & 0.0113   & 0.81 &  405   & 25 \\
 &     & N & K00719.a & $15.7687_{-6}^{+5}$  &   $966.435_{-8}^{+10}$ & $0.0059_{-10}^{+6}$   & 0.42 &  12   & 9 \\
 &     & I & K00719.I & $6.2\pm0.6$    & -   &  -  & 0.6 & -    &-  \\
 &  0.49   & I & K00719.II & $14\pm2$   &  -  &  -  & 0.7 &   -  & - \\
 &     & I & K00719.III & $20\pm2$    & -   &  -  & 0.8 & -    &-  \\
 &     & E & K00719.IV & $66\pm7$  & -   & -  & 1.1 &   -  & - \\

10271806&0.22 & O &K00733.01&	5.925020&	1002.714 & 0.01071 & 2.84 & 328 & 64 \\
        & & O &K00733.02&		11.34930&	967.318  & 0.01093 & 2.47 & 195 & 24 \\
        & & O &K00733.03&		3.132940&	968.677  & 0.00878 & 1.52 & 60 & 25 \\
        & & O &K00733.04&		18.64390&	974.620  & 0.00993 & 2.55 & 13  & 14\\
        &4.06 &N&K00733.a \tablenotemark{c}&	$15.11133_{-4}^{+3}$	&	$972.676_{-2}^{+2}$  & $0.0113_{-4}^{+6}$ & 3.00 & 13  & 8 \\
        &    & E  &K00733.I \tablenotemark{d}& $36\pm4$    &	 -        & - &  2.8 &- &- \\
\hline\\
4077526 & 1.06 & O & K01336.01&	10.218500&	969.871 & 0.02150 &2.6 & 329 & 19 \\
        & &O & K01336.02&	15.573800&	965.438 & 0.02169 &2.6  & 177 & 11 \\
        & &O &K01336.03&	40.101000&	965.715 & 0.02260 &2.7  & 18 &13 \\
        & &O &K01336.04&	4.458250 &	967.387 & 0.01431 &1.7  & 31 & 9\\
        & 1.35 &N& K01336.a  &    $27.5066_{-1}^{+9}$ &  $977.882_{-3}^{+3} $ & $0.0086_{-5}^{+6}$ &1.04 & nan & nan\tablenotemark{e}\\
        & & I & K01336.I &   $26\pm 3$ &  -      &    -  &  2.4  & -& -\\
        &0.19 & I & K01336.II &   $6.8\pm 3$ &  -      &    -  &  1.7  &- &- \\
        & &E & K01336.III &   $61\pm 6$ &  -      &    -  &  3.0 &- &-\\
\hline\\
7747425 &3.26 &O &K01952.01&	8.010350&	1002.714 & 0.01692 & 1.87 & 161 & 27 \\
        & &O&K01952.02&	27.665000&	976.876  & 0.01867 & 2.06 & 14 & 17 \\
        & &O&K01952.03&	5.195500&	968.677  & 0.01135 & 1.25 & 217 & 15\\
        & &O&K01952.04&	42.469900&	979.305  & 0.01821 & 2.01 & 30 & 13\\
        & 0.64 & N &K01952.a&	$13.27242_{-1.3}^{+9} $  &  $973.014_{-4}^{+4}$ & $0.0077_{-5}^{+9}$ & 0.85 & 14& 9 \\
        & &I & K01952.I&	$13\pm 2 $ &- & - & 1.5 &- &-\\
        & 0.01 &I&K01952.II&	$19\pm 2 $ & -& - & 1.6 &- &-\\
        &  &E &K01952.III&	$65\pm 7 $ &- &-  & 2.2 &- &-\\
\hline\\
7673192  &0.98 & O&K02722.01&	6.124820  & 967.935 & 0.01093 & 1.47 & 215 & 28 \\
        & &O&K02722.02&		11.242800 & 969.322  & 0.01071 & 1.44 & 184 & 17\\
        & &O&K02722.03&		4.028710  & 966.775 & 0.00878 & 1.18 & 17 & 15 \\
        & &O&K02722.04&		8.921080  & 970.494  & 0.00993 & 1.33 & 142 & 16\\
        & &N&K02722.a& $16.5339_{-1}^{+1}$ &	$968.407_{-5}^{+7}$  & $0.0086_{-5}^{+8}$ & 1.16 & 31 & 13 \\
       &  &E&K02722.I& $16.8\pm1.0$    &	  -       & - &  2.8 &- &- \\
\hline\\
10397751 &1.69 &O&K02859.01 &3.446219& 965.2274&0.0132&1.27 & 89 & 16 \\ 
        & &O&K02859.02 &2.005396&966.2442 &0.0074&0.72 &99 &13 \\
        & &O&K02859.03 &4.288814&965.3824 &0.0075&0.72 &105 &11\\
        & &O&K02859.04 &2.905106&965.5643&0.0086&0.83 &150 &10\\
        &1.07 &N&K02859.a &$5.43105_{-4}^{+6}$ &$967.41401_{-3}^{+5}$&$0.0079_{-6}^{+10}$ &0.76 & 160 & 10 \\
        & &I &K02859.I &$2.41\pm0.1$    &    -          & - &0.6 &- & -\\
        & &E &K02859.II & $5.2\pm0.3$&-&-&0.8 & -&-\\

\enddata
\tablenotetext{a}{The first value for every system is based the fitting 
before the insertion; the value next to the detected planet is based on the 
fitting including the detection; the value next to the predicted planet is 
based on the fitting including the prediction (but exclude the detection). 
A smaller $\chi^2_\nu$ indicates tighter fitting to the TB relation.}
\tablenotetext{b}{``O" indicates KOI candidate from {\em Kepler} team. ``N" 
indicates new detections from this work. ``I" indicates inserted 
planets (predicted). ``E" indicates extrapolated planet (predicted).}
\tablenotetext{c}{KOI number with letter indicates this is a new detection.}
\tablenotetext{d}{KOI number with roman numbers indicates this is a 
predicted planet.}
\tablenotetext{e}{This candidate is found by identifying the BLS peaks visually 
around the predicted periods.
}
\end{deluxetable}
\end{landscape}
\clearpage

\begin{deluxetable}{cccc}
\centering
\tablewidth{0pc}
\tablecaption{Detection Summary \label{tab:expect}}
\tablehead{
\colhead{Catagory\tablenotemark{a}} & \colhead{Num. Predictions \tablenotemark{b}} & \colhead{Num. Expectations \tablenotemark{c}} & \colhead{Num. Detections} \tablenotemark{d}
}
\startdata
SI & 10 (39)  & $7.8_{-1.7}^{+3.5}$  & 4 (4)  \\
CI & 8  (20)  & $4.2_{-0.4}^{+1.2}$  & 0 (0)  \\
CE & 17 (17)  & $2.2_{-0.6}^{+1.8}$  & 1 (1)  \\
SE & 21 (21)  & $0.65_{-0.2}^{+0.8}$  & 0 (0)  
\enddata
\tablenotetext{a}{``S" (C) - the systems are sparse (complete); 
``I"(E)-the systems need (do not need) insertion. For detailed definition 
of each category, see Section \S2.}
\tablenotetext{b}{The first number indicates the number of systems predicted, 
the second number indicates the total number of planets predicted in these 
systems.}
\tablenotetext{c}{The expected number of planets to be detected after 
accounting observational bias. We estimate this using the most likely 
Rayleigh distribution for inclination with 
$\sigma_i = 1.8_{-0.8}^{+0.5}\,^{\circ}$ and the turn-over power-law for size 
correction. The upper bound and lower bound are $3-\sigma$ limitations.}
\tablenotetext{d}{The detected number of systems (planets) that match the 
predictions.}
\end{deluxetable}
\clearpage
\section{Acknowledgements}
We thank the anonymous referee for the insightful comments. 
We thank Zhou, G. and Petrovich, C. for their comments on the draft and helpful discussions. 
Work by XCH and GB were supported by the NASA NNX13AJ15G and NSF AST1108638 grants.


\begin{thebibliography}{}
\bibitem[\protect\citeauthoryear{Borucki et al.}{2010}]{Borucki:2010} Borucki, W.~J., Koch, D., Basri, G., et al.\ 2010, Science, 327, 977 
\bibitem[\protect\citeauthoryear{Barclay et al.}{2013}]{Barclay:2013} Barclay, T., Rowe, J.~F., Lissauer, J.~J., et al.\ 2013, Nature, 494, 452
\bibitem[\protect\citeauthoryear{Batalha et al.}{2013}]{Batalha:2013} Batalha, N.~M., Rowe, J.~F., Bryson, S.~T., et al.\ 2013, ApJS, 204, 24
\bibitem[\protect\citeauthoryear{Bovaird \& Lineweaver}{2013}]{Bovaird:2013} Bovaird, T., \& Lineweaver, C.~H.\ 2013, arXiv:1304.3341
\bibitem[\protect\citeauthoryear{Ciardi et al.}{2013}]{Ciardi:2013} Ciardi, 
D.~R., Fabrycky, D.~C., Ford, E.~B., et al.\ 2013, ApJ, 763, 41 
\bibitem[\protect\citeauthoryear{Chambers et al.}{1996}]{Chambers:1996} Chambers, J.~E., Wetherill, G.~W., \& Boss, A.~P.\ 1996, Icarus, 119, 261 
\bibitem[\protect\citeauthoryear{Dong \& Zhu}{2012}]{Dong:2012} Dong, S., \& Zhu, Z.\ 2012, arXiv:1212.4853 
\bibitem[\protect\citeauthoryear{Fabrycky et al.}{2012}]{Fabrycky:2012} Fabrycky, D.~C., Lissauer, J.~J., Ragozzine, D., et al.\ 2012, arXiv:1202.6328
\bibitem[\protect\citeauthoryear{Hayes \& Tremaine}{1998}]{Hayes:1998} Hayes, W., \& Tremaine, S.\ 1998, Icarus, 135, 549
\bibitem[\protect\citeauthoryear{Hansen \& Murray}{2013}]{Hansen:2013} Hansen, B., \& Murray, N.\ 2013, arXiv:1301.7431 
\bibitem[\protect\citeauthoryear{Hills}{1970}]{Hills:1970} Hills, J.~G.\ 1970, BAAS, 2, 199
\bibitem[\protect\citeauthoryear{Huang et al.}{2013}]{Huang:2013} Huang, X., Bakos, G.~{\'A}., \& Hartman, J.~D.\ 2013, MNRAS, 429, 2001
\bibitem[\protect\citeauthoryear{Howard et al.}{2012}]{Howard:2012} Howard, A.~W., Marcy, G.~W., Bryson, S.~T., et al.\ 2012, ApJS, 201, 15
\bibitem[\protect\citeauthoryear{Latham et al.}{2011}]{Latham:2011} Latham, D.~W., Rowe, J.~F., Quinn, S.~N., et al.\ 2011, ApJL, 732, L24
\bibitem[\protect\citeauthoryear{Laskar}{2000}]{Laskar:2000} Laskar, J.\ 2000, Physical Review Letters, 84, 3240
\bibitem[\protect\citeauthoryear{Lissauer et al.}{2011}]{Lissauer:2011} Lissauer, J.~J., Ragozzine, D., Fabrycky, D.~C., et al.\ 2011, ApJS, 197, 8
\bibitem[\protect\citeauthoryear{Lissauer et al.}{2012}]{Lissauer:2012} Lissauer, J.~J., Marcy, G.~W., Rowe, J.~F., et al.\ 2012, ApJ, 750, 112 
\bibitem[\protect\citeauthoryear{Li et al.}{1995}]{Li:1995} Li, X.~Q., Zhang, H., \& Li, Q.~B.\ 1995, A\&A, 304, 617
\bibitem[\protect\citeauthoryear{Petigura et al.}{2013a}]{Petigura:2013a} Petigura, E.~A., Marcy, G.~W., \& Howard, A.~W.\ 2013, ApJ, 770, 69
\bibitem[\protect\citeauthoryear{Petigura et al.}{2013b}]{Petigura:2013b} Petigura, E.~A., 
Howard, A.~W., \& Marcy, G.~W.\ 2013, arXiv:1311.6806 
\bibitem[\protect\citeauthoryear{Patterson}{1987}]{Patterson:1987} Patterson, C.~W.\ 1987, Icarus, 70, 319
\bibitem[\protect\citeauthoryear{Petrovich et al.}{2013}]{Petrovich:2013} Petrovich, C., Malhotra, R., \& Tremaine, S.\ 2013, ApJ, 770, 24
\bibitem[\protect\citeauthoryear{Steffen et al.}{2012a}]{Steffen:2012a} Steffen, J.~H., Fabrycky, D.~C., Ford, E.~B., et al.\ 2012, MNRAS, 421, 2342 
\bibitem[\protect\citeauthoryear{Steffen et al.}{2012b}]{Steffen:2012b} Steffen, J.~H., Ford, E.~B., Rowe, J.~F., et al.\ 2012, ApJ, 756, 186 
\bibitem[\protect\citeauthoryear{Steffen}{2013}]{Steffen:2013} Steffen, J.~H.\ 2013, arXiv:1301.2394
\end{thebibliography}
\end{document}